\documentclass[10pt,a4paper,twocolumn]{article}
\pdfoutput=1
\usepackage[utf8]{inputenc}
\usepackage[english]{babel}
\usepackage[T1]{fontenc}
\usepackage{textcomp}
\usepackage{amsmath}
\usepackage{amsthm}
\usepackage{amsfonts}
\usepackage{amssymb}
\usepackage{makeidx}
\usepackage{graphicx}
\usepackage{mdwlist}
\usepackage{times}
\usepackage[retainorgcmds]{IEEEtrantools}
\usepackage{fullpage}
\usepackage{color}
\usepackage[bottom]{footmisc}

\title{Van Wijngaarden grammars,\\
metamorphism and \emph{K}-ary malwares.}
\author{Gueguen Geoffroy\\
ESIEA Laval\\
Laboratoire de cryptologie\\
et de virologie opérationelles ($C + V)^{O}$,\\
38 rue des Dr Calmette et Guérin\\
53000 Laval, France\\
\texttt{geoffroy.gueguen@esiea-ouest.fr}}
\date{\today}

\newtheorem{definition}{Definition}

\begin{document}
\maketitle

\begin{abstract}
\emph{Grammars are  used to describe sentences structure, thanks to some sets of rules, which depends on the grammar type. A classification of grammars has been made by Noam Chomsky, which led to four well-known types. Yet, there are other types of grammars, which do not exactly fit in Chomsky's classification, such as the two-level grammars. As their name suggests it, the main idea behind these grammars is that they are composed of two grammars.}

\emph{Van Wijngaarden grammars, particularly, are such grammars. They are interesting by their power (expressiveness), which can be the same, under some hypotheses, as the most powerful grammars of Chomsky's classification, i.e.\ Type 0 grammars. Another point of interest is their relative conciseness and readability.}

\emph{Van Wijngaarden grammars can describe static and dynamic semantic of a language. So, by using them as a generative engine, it is possible to generate a possibly infinite set of words, while assuring us that they all have the same semantic. Moreover, they can describe K-ary codes, by describing the semantic of each components of a code.}
\end{abstract}

\section{Introduction}

Grammars are mostly used to describe languages, like programming languages, in order to parse them. In this paper, we are not interested in the parsing problem of a language. On the contrary, the objective is to use a grammar from which the word parsing problem is known to be hard (as in NP), or even better, undecidable. Indeed, if one wants to do some metamorphism through the use of a grammar, one may want to avoid grammars for which techniques to build practical word recognizers of a language are known.

Van Wijngaarden grammars are different than the one which fall in Chomsky\rq{}s classification. Their writing is particular, and above all, their production process is quite different than the grammars in Chomsky\rq{}s hierarchy. We will see that these grammars may be used as \lq\lq{}code translators\rq\rq{}. They indeed have some rules which allow them to be very expressive.

\section{Metamorphism vs. Polymorphism}

The difference between polymorphism and metamorphism is often not very clear in people's mind, so we describe it quickly in this section.

\subsection{Polymorphism}
\begin{figure}[h]
 \begin{center}
  \includegraphics[width=0.4\textwidth]{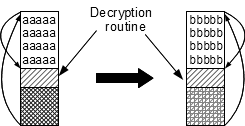}
  \caption{Two files infected by the same virus.}
  \label{polymorphism}
 \end{center}
\end{figure}

\begin{figure}[h]
 \begin{center}
  \includegraphics[width=0.4\textwidth]{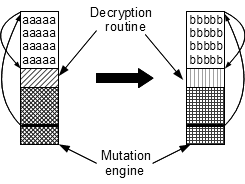}
  \caption {Two files infected by the same virus.}
  \label{polymorphism2}
 \end{center}
\end{figure}

Polymorphism first appeared to counter the detection scheme of AV companies which was, and still is for a main part, based on signature matching. The aim of virus writers was to write a virus whose signature would change each time it evolves. In order to do so, the virus body is encrypted by an encryption function and it is decrypted by its decryptor at the runtime. The key used to encrypt each copy of the virus is changed, so that each copy has a different body (Figure \ref{polymorphism}).
Another technique that can be used is to apply a different encryption scheme for each copy of the code. Of course such a technique alone is not enough to evade signature detection as it only shifts the problem, the decryptor being a good candidate for a signature. To resolve this, the decryption routine has to be changed too between each copy of the virus. To do so, virus writers include a mutation engine, which is also encrypted during the propagation process, and which is used to randomly generate a new decryption routine so it is different from copy to copy (\makebox{Figure \ref{polymorphism2}}).
While in the first case the decryption routine can be used as a signature, this is not the case in the second one. Indeed, the decryption routine changes from mutation to mutation, thanks to the engine. \\
The mutation engine cannot be used as a signature neither, because it is a part of the body, thus it is encrypted. The propagation process can be summed up in five \makebox{steps :}
\begin{itemize}
\item The decryption routine decrypts the encrypted body;
\item The body is executed;
\item The code calls the mutation engine (which is decrypted at this stage) to transform the decryption routine;
\item The code and the mutation engine are encrypted;
\item The transformed decryption routine and the new encrypted body are then appended onto a new program.
\end{itemize}

\subsection{Metamorphism}

Metamorphism differs from polymorphism in the fact that there is no use of a decryption routine, because there is no encryption process. In other words, while a polymorphic code has to decrypt itself before it can be executed, a metamorphic one is executed directly.

\begin{figure}[h]
 \begin{center}
  \includegraphics[width=0.35\textwidth]{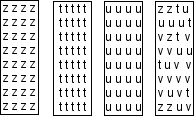}
  \caption{Four equivalent codes.}
  \label{metamorphism}
 \end{center}
\end{figure}

Indeed, a metamorphic engine can be seen as a \lq\lq{}semantic translator\rq\rq{}. The idea is to rewrite a given code into another syntactically different, yet semantically equivalent one (Figure \ref{metamorphism}).
Different techniques can be used to build an efficient metamorphic engine. Among these techniques we can observe :
\begin{itemize}
\item Junk code insertion :
a junk code is a code that is useless for the main code to perform its task.
\item Variable renaming :
the variables used between different versions of the code are different.
\item Control flow modifications :
some instructions are independent from each other, and so, can be swapped. Otherwise instructions can be shuffled and linked by jumps.
\end{itemize}

\section{Grammars}

In this section, we recall what formal grammars are and the link they have with languages.

\subsection{What is a grammar}

\begin{definition}
Let $\Sigma$ be a finite set of symbols called alphabet. A formal grammar G is defined by the \makebox{4-tuple} \makebox{$G = (V{}_{N}, V{}_{T}, S, P)$} \makebox{where :}
\begin{itemize}
\setlength{\itemsep}{0 pt}
\item $V_{N}$ is a finite set of \emph{non-terminal} symbols,  \makebox{$V_{N} \cap \Sigma^{\ast} = \emptyset$};
\item $V_{T}$ is a finite set of \emph{terminal} symbols, \makebox{$V{}_{N} \cap V{}_{T} = \emptyset$};
\item S $\in V{}_{N}$ is the \emph{starting symbol} of the grammar;
\item P $\subseteq (V{}_{T}\cup V{}_{N})^{\ast}\times(V{}_{T}\cup V{}_{N})^{\ast}$ is a set of \emph{production rules}.
\end{itemize}
\end{definition}

Basically, a grammar can be seen as a set of rewriting rules over an alphabet. An alphabet is a finite set of symbols (like \lq{}$a$\rq{}, \lq{}$b$\rq{}). We distinguish two sets of symbols. The first one is the set of \emph{non-terminal} symbols, and the second is the set of \emph{terminal} symbols. Non-terminal symbols are symbols which are used to be replaced by the right-hand side of a production rule. On the contrary, terminal symbols are symbols which cannot be modified by a rule. Of course, the two sets are disjoint. A rewriting rule is a rule which defines how a given sequence of symbols can be rewritten into another sequence of symbols. A special symbol, called the start symbol, is used to specify where the rewriting must start. This particular symbol belongs to the set of non-terminal symbols. We then note $G = (N,T,S,P)$ to define the grammar $G$ composed of the set of non-terminal symbols $N$, the set of terminal symbols $T$, the starting symbol $S$, and the set of production (rewriting) rules $P$. 

\begin{definition}
Let $G = (V{}_{N}, V{}_{T}, S, P)$ be a formal grammar. The language described by G is L(G) = $\lbrace x \in \Sigma^{\ast} \mid S \rightarrow^{\ast} x \rbrace$ .
\end{definition}

Grammars are used to describe languages. A language is a set of words, each word being a sequence of symbols. A word may or may not have a meaning nor a structure. For instance, the grammar \makebox{$G$ $=$ $(\lbrace S\rbrace$, $\lbrace a\rbrace$, $S$, $\lbrace S \rightarrow aS$; $S \rightarrow a\rbrace)$} (here \makebox{$N = \lbrace S\rbrace$}, \makebox{$T = \lbrace a\rbrace$}, \makebox{$S = S$}, and \makebox{$P = \lbrace S \rightarrow aS; S \rightarrow a\rbrace)$} describes the language \makebox{$L(G) = \lbrace a^{n} \mid n \geq 1 \rbrace$} (i.e.\ the words \lq{}$a$\rq{}, \lq{}$aa$\rq{}, \lq{}$aaa$\rq{}, $\ldots$). There exist different forms which are used to represent grammars. For convenience, we will write the production rules of a grammar as \makebox{follows :}
\begin{IEEEeqnarray*}{s}
S $\rightarrow$ aS \\
S $\rightarrow$ a
\end{IEEEeqnarray*}
When some rules share the same left-hand side, as it is the case here, we can shrink the different alternatives in one rule, separated by a \lq{}$\mid$\rq{} :
\begin{IEEEeqnarray*}{s}
S $\rightarrow$ aS $\mid$ a
\end{IEEEeqnarray*}

To generate a word from these rules one proceeds as follows : start from the starting symbol and replace it by one of its alternatives. Then two cases have to be considered :
\begin{itemize}
\item[-] either a sequence of symbols of the produced sentential form matches the left-hand side of a rule;
\item[-] either it is not the case and, if the sentential form does not contain any non-terminal symbols, it is a word of the language described by the grammar.
\end{itemize}
Whenever a sequence of symbols matches the left-hand side of a rule, it is replaced by one of the alternatives of the rule, and the process goes on until no more match is found.

As an example, take the above rule. The starting word is \lq{}$S$\rq{}. Suppose that \lq{}$S$\rq{} produces the sentential form \lq{}$a$\rq{}. As \lq{}$a$\rq{} does not match any left-hand side of the rules at our disposal, and as it is a terminal symbol, it is a word of the language. Now suppose that \lq{}$S$\rq{} produces the sentential form \lq{}$aS$\rq{}. The non-terminal \lq{}$S$\rq{} in \lq{}$aS$\rq{} matches the left-hand side of one of the rules, so we replace it by one of its alternatives : \lq{}$a$\rq{} or \lq{}$aS$\rq{}. We thus obtain the sentential forms \lq{}$aa$\rq{} or \lq{}$aaS$\rq{}. Hence, the words generated by the above rule are : $a$, $aa$, $aaa$, \makebox{$aaaa$, $\ldots$}

Now, if we use some x86 instructions as the terminal symbols, we can write rules which will generate x86 instructions sequences \cite{Filiol07-2,DBLP:journals/virology/Zbitskiy09}. From a given sequence of instructions, it is easy to write a grammar which will generate it. For instance, the instruction sequence :
\begin{IEEEeqnarray*}{s}
mov eax, key\\
xor [ ebx ], eax\\
inc ebx
\end{IEEEeqnarray*}

can be generated by the following production rules :
\begin{IEEEeqnarray*}{s}
$S \rightarrow$ mov eax, key $T$\\
$T \rightarrow$ xor [ ebx ], eax $U$\\
$U \rightarrow$ inc ebx $V$
\end{IEEEeqnarray*}
The instruction sequence is thus represented by the sequence of non-terminal symbols S $\rightarrow$ T $\rightarrow$ U $\rightarrow$ V, the non-terminal S being rewritten into the sentential form \lq\lq{}mov eax, key $T$\rq\rq{}, which is then rewritten into the sentential form \lq\lq{}mov eax, key\ \ xor [ ebx ], eax $U$\rq\rq{}, etc\ldots

Once the production rules are defined, one may want to generate an equivalent sequence of instructions. It is rather easy :
\begin{IEEEeqnarray*}{ss}
$S \rightarrow$ & mov eax, key $T$ $\mid$ push key; pop eax $T$\\
$T \rightarrow$ & xor [ ebx ], eax $U$ $\mid$ mov ecx, [ ebx ]; \\
& and ecx, eax; not ecx; or [ ebx ], eax; \\ & and [ ebx ], ecx $U$ \\
$U \rightarrow$ & inc ebx $V$ $\mid$ add ebx, 1 $V$
\end{IEEEeqnarray*}
The production rules now generate 8 $(2\times2\times2)$ different sequences, each of them acting the same. In a same manner, one may want to add some junk code. This can be done by adding a new non-terminal which generates \lq\lq{}useless\rq\rq{} instructions\footnote{Care must be taken on the place where to add these instructions, as they may modify some flags which are check later, e.g.\ by a $jcc$ instruction.} :
\begin{IEEEeqnarray*}{ss}
$S \rightarrow$ & $G$ mov eax, key $T$ $\mid$ $G$ push key; pop eax $T$\\
 & \ldots \\
$G \rightarrow$ &  add edx, 1; dec edx $\mid$ push eax; add esp, 4
\end{IEEEeqnarray*}

For this example, the addition of the rule $G$, which is composed of only two alternatives, increases the number of instruction sequences that can be generated to 216 $(6\times6\times6)$. This number can be made infinite pretty easily, by adding alternatives which generate only junk code for example, like :
\begin{IEEEeqnarray*}{ss}
$S \rightarrow$ & $G$ $S$ $\mid$ mov eax, key $T$ $\mid$ push key; pop eax $T$\\
& or else \\
$G \rightarrow$ & $G$ $G$ $\mid$  add edx, 1; dec edx $\mid$ push eax; add esp, 4
\end{IEEEeqnarray*}
\subsection{Classification of grammars}

Chomsky provided a well-known classification of grammars \cite{Cho56}. He defined four main types, from \makebox{\emph{Type 0}} to \emph{Type 3}, each type defining a set of languages, each of them being a subset of the set described by any lower numbered grammar. In other words, \emph{Type 0} are the most general grammars, while \emph{Type 3} are the most restrictive. Among these grammars, \emph{Type 2}, also called context-free grammars, are the most popular. They describe context-free languages. Most of the programming languages are described by such grammars. The rules of \emph{Type 2} grammars have the following \makebox{form :}
\begin{center}
U $\rightarrow$ V
\end{center}
Where U is a single non-terminal symbol, and V belongs to $(N\times T)^{\ast}$.

In other words, U can be rewritten as a possibly empty sequence of terminal and non-terminal symbols. The name context-free comes from the fact that the left-hand side of a rewriting rule is a single non-terminal, so the rewriting does not depend of what may be next to it in a sentential form, unlike in \emph{Type 0} and \emph{Type 1} grammars.
We have the relation \emph{Type 0} $\supset$ \emph{Type 1} $\supset$ \emph{Type 2} $\supset$ \emph{Type 3}. Thus, \emph{Type 0} grammars can define all the languages that are definable by \emph{Type 1, Type 2} or \emph{Type 3} grammars.

\section{Van Wijngaarden grammars}
\subsection{Context-sensitivity restrictions}

Context-sensitive languages are more complex than context free languages because one part of the string may \lq\lq{}interact\rq\rq{} with the structure of the other parts of the string. Once a non-terminal symbol has been produced in a sentential form in a context-free grammar, its further development is independent of the rest of the sentential form, whereas a non-terminal symbol in a sentential form of a context-sensitive grammar has to look at its neighbours, on its left and on its right, to see what are the production rules that are allowed for it. So a context-free grammar cannot express some \lq\lq{}long-range\rq\rq{} relations.

Yet, these relations are often valuable, as they make possible some fundamental properties of words to be described (like the only use of variables that have been declared). Programming languages are usually context-sensitive. For example a user is usually not allowed to use a variable that has not been created. So as it is not possible to express such properties through a context-free grammar, a solution, which is used most of the time, is to describe the structure of the correct words by a context-free grammar. The properties are checked by a separate program after that the word has been recognize by the grammar (though it may not belong to the \lq\lq{}real\rq\rq{} language). However, this solution is not very satisfactory as the interest of using a grammar is to have a (formal) description of all the properties of the language.

One can ask why a context-sensitive grammar is not used to describe the language. Actually this would pose some problems. Indeed, in general, context-sensitive languages cannot be parsed efficiently. Moreover, even though context-sensitive grammars have the power to express some long-ranged relations in a sentential form, they don't do it in a way that is easily understandable.

Also it would make sense that after having written a grammar for $a^{n}b^{n}c^{n}$, the writing of $a^{n}b^{n}c^{n}d^{n}$ would work the same way. But this is not the case : the grammar for $a^{n}b^{n}c^{n}d^{n}$ is more complex. The reason behind that is that to express a long-range relation, informations have to flow through the sentential form, thanks to the non-terminal symbols (which look at their neighbours to rewrite a sentential form into another). Thus it requires almost all rules to know something about almost all the other rules.

Several grammar forms which make these relations more readable and easier to construct have been created. Among them are Van Wijngaarden grammars.

\subsection{VW grammar definition}

Basically, a VW grammar can be seen as the composition of two context-free grammars (that is why such grammars are also called two-level grammars). The first context-free grammar is used to generate a set of terminal symbols which will act as non-terminals for the second context-free grammar.

Before going further, a few terms have to be introduced.
\begin{itemize}
\item A $protonotion$ is a sequence of small syntactic marks ;
\item A $metanotion$ is a sequence of big syntactic marks which is defined in a metarule ;
\item A $hypernotion$ is a possibly empty sequence of metanotions and protonotions ;
\item A $metarule$ defines a metanotion as a possibly empty sequence of hypernotions ;
\item A $hyperrule$ defines a sequence of hypernotions as another sequence of hypernotions, separated by a comma. Actually, they represent a possibly infinite set of production \makebox{rules ;}
\item A VW grammar is defined by a set of metarules (or metaproduction rules) and a set of hyperrules ;
\item Whenever a metanotion appears more than once in a hyperrule, each of its occurrence have to be replaced consistently throughout the rule. This is called the \emph{Uniform Replacement Rule}.
\end{itemize}

\begin{definition}\cite{1781176}
A Van Wijngaarden grammar is a grammar G = ( M, V, N, T, $R_{M}$, $R_{V}$, S ) \makebox{with :}
\begin{IEEEeqnarray*}{sts}
$M$ & : & a finite set of metanotions  \\
$V$ & : & a finite set of metaterminals, $M \cap V = \emptyset$ \\
$N$ & : & a finite set of hypernotions, $N \subseteq ( M\cap V)^{+} )$ \\
$T$ & : & a finite set of terminals \\
$R_{M}$ & : & a finite set of metarules, $X \rightarrow Y$ with \\
&& $X \in M$, $Y \in ( M\cap V)^{\ast}$ such that for all \\ && $W \in M$, $(M, V, W, R_{M})$ is a context-free \\ && grammar \\
$R_{V}$ & : & a finite set of hyperrules \\
\IEEEeqnarraymulticol{3}{s}{$S \in N$ : the starting symbol }
\end{IEEEeqnarray*} 
\end{definition}
The first set of rules are the metarules. They represent a modified grammar in which the non-terminals are replaced by metanotions, and the terminals are replaced by protonotions. The second set of rules are the hyperrules. They represent some possibly infinite set of production rules.

In order to make a distinction between the metarules, the hyperrules, and the production rules, the production symbol is changed. Instead of the symbol \lq{}$\rightarrow$\rq{} we use \lq{}$::$\rq{} for the metarules and \lq{}$:$\rq{} for the hyperrules. To separate the different alternatives of a rule, the symbol \lq{}$;$\rq{} is used instead of \lq{}$\mid$\rq{}. In metarules, members are separated by a blank, and in hyperrules, by a comma. The metanotions have to be chosen wisely, so that any sequence of metanotions is not also a different sequence of metanotions. For instance, if we have a metanotion \emph{X} and a metanotion \emph{Y}, then the metanotion \emph{XY} should be avoided as it would induce some ambiguity.

To make it clearer, here is a VW grammar which describes the language \makebox{$L = \{ a^n b^n c^n\mid n >= 1 \}$} (i.e.\ $abc$, $aabbcc$, $aaabbbccc$, \ldots) :
\begin{IEEEeqnarray*}{us}
N &:: i N; i.\\
A &:: a; b; c. \medskip \\
S & : aN, bN, cN.\\
AiN & :  A symbol, AN.\\
Ai & :  A symbol.
\end{IEEEeqnarray*}
The first two rules are the metarules, and the last three are the hyperrules. The metanotions are $N$ and $A$. The hypernotions are $AiN$, $Ai$, $A$, $AN$, $aN$, $bN$, and $cN$.

In the definition of a VW grammar, a member is a terminal symbol if it ends in $symbol$ (like \lq{}b symbol\rq{} for the terminal symbol \lq{}b\rq{}), otherwise it is a non-terminal. So, here the rule \lq\lq{}Ai : A symbol.\rq\rq{} produces the terminal symbols $a, b$ and $c$.

The metanotion $N$ produces an infinite set of $i$. The $i$'s act as a counter for the number of letters to be produced. Indeed, as we said, the hypernotions describe a possibly infinite set of production rules. For instance here, the rule \lq\lq{}AiN : A symbol, AN.\rq\rq{} actually produces the \makebox{rules :}
\begin{IEEEeqnarray*}{us}
aii & : a symbol, ai. \\
aiii & : a symbol, aii. \\
\vspace{5pt}
& \hspace{15pt} etc$\ldots$ \\
bii & : b symbol, bi.\\
biii & : b symbol, bii.\\
\vspace{5pt}
& \hspace{15pt} etc$\ldots$ \\
cii & : c symbol, ci.\\
ciii & : c symbol, cii.\\
& \hspace{15pt} etc$\ldots$
\end{IEEEeqnarray*}
To obtain these sets, the metanotion $A$ is replaced consistently by all the words it can generate. Here these are $a$, $b$ and $c$. So we obtain the following three rules :
\begin{IEEEeqnarray*}{us}
aiN & : a symbol, aN.\\
biN & : b symbol, bN.\\
ciN & : c symbol, cN.
\end{IEEEeqnarray*}
Then the same thing is done with the metanotion N. As it generates the infinite language \makebox{$L(N) = \lbrace i^{n} \mid n \geq 1 \rbrace$} (i.e. \lq{}$i$\rq{}, \lq{}$ii$\rq{}, \lq{}$iii$\rq{}...), we obtain the above three sets of infinite production rules.

\subsection{Place in Chomsky's hierarchy}

By construction, Van Wijngaarden grammars do not belong to any category of Chomsky\rq{s} classification. However, one can compare the expressive power of a Van Wijngaarden grammar and the different types of Chomsky\rq{}s hierarchy. In terms of expressive power, they are in fact equivalent to \emph{Type 0} grammars. In a sense, they are even more powerful than \emph{Type 0} grammars since they can handle infinite symbols sets. For instance, as shown in Figure \ref{infiniteset}, a Van Wijngaarden grammar can produce the \makebox{set :}
\begin{center}$S\ =\ \lbrace\ t_{1}^{n}\cdots t_{k}^{n} \mid n \geq 0,\ k > 0,\ t_{1} \cdots t_{k}$ are different symbols $\rbrace$\end{center}
A \emph{Type 0} grammar cannot generate this set since its number of (terminal) symbols is infinite.
\begin{figure}[h]
\begin{IEEEeqnarray*}{us}
N & :: n N; $\varepsilon$. \\
C & :: i ; i C. \\
\\
S & : N\ i tail. \\
N C tail\ & : N, C, N C i tail ; $\varepsilon$. \\
N n C & : C symbol, N C. \\
C & : $\varepsilon$.
\end{IEEEeqnarray*}
\caption{A grammar handling an infinite alphabet}
\label{infiniteset}
\end{figure}

Sintzoff \cite{sintzoff} showed that there exists a Van Wijngaarden grammar for every semi-thue system\footnote{A semi-thue system is a string rewriting system. It is equivalent to Chomsky\rq{}s \emph{Type 0} grammars.}. Van Wijngaarden \cite{681877} showed that a Van Wijngaarden grammar can simulate a \emph{Turing Machine}. Thus, both proved that these grammars are at least as powerful as \emph{Type 0} grammars (i.e.\ that they are Turing complete). As a consequence, parsing of these grammars is undecidable in general.
On a side note, it is to be noted that, if the first set of rules, i.e.\ the metarules, does not generate an infinite language, then the Van Wijngaarden grammar is equivalent to a standard context-free grammar. 
Indeed, if the language generated by a metarule is finite, one can write as much production rules as there is words in the language, and the consistent
substitution can be \lq\lq{}emulated\rq\rq{} by the addition of rules which produce only one sentence. For instance the grammar :
\begin{IEEEeqnarray*}{us}
S $\rightarrow$ & P1 BODY P2 $\mid$ P3 BODY P4\\
P1 $\rightarrow$ & (\\
P2 $\rightarrow$ & )\\
P3 $\rightarrow$ & <\\
P4 $\rightarrow$ & >
\end{IEEEeqnarray*}
ensures that the opening bracket matches the ending one. By increasing the number of rules of the grammar, we can express more and more context-sensitive conditions. It follows that if we have an infinite collection of context-free grammar rules, we can express any number of context-sensitive conditions, and so we can achieve full context-sensitivity. As said in the beginning of this section, this is the idea behind Van Wijngaarden grammars : a VW grammar can be seen as the composition of two context-free grammars. The first context-free grammar is used to generate a language which can in turn be described by the second context-free grammar. Nonetheless, as mentioned in the previous section, it is possible to produce every words of the languages they may describe.

\subsection{VW grammars and word generation}

Dick Grune \cite{2356} made a program which can produce all the sentences of a Van Wijngaarden grammar. The program reads a grammar on its input, and then the generation of the words starts. If the input\rq{s} grammar describes an infinite language, then an infinite number of words will be produced. We modified some parts of this program in order to implement our mutation engine, and we have written a VW grammar based on the x86 instructions set.

It is not possible to generate the words of a Van Wijngaarden grammar in the same way that those of a context-free grammar are. Indeed, to generate a terminal production for a context-free language, we start from the start symbol. Intermediate results of a production (sentential forms) are stored in a queue. To rewrite a sentential form, we consider initially the first sentential form in the queue. Then, we search for a sequence of symbols which match the left-hand side of a production rule. If such a match is found, the sentential form is replaced by all its alternatives by making as much copies as the number of alternatives, and each copy is appended at the end of the queue. If no match is found, it means the sentential form is a terminal production.

This process cannot be applied to Van Wijngaarden grammars, as there may be an infinite number of left-hand side resulting from a same hyperrule. Actually, it would require us to scan all the possible left-hand side of the hyperrule, so you may have to look at an infinite number of left-hand side to know if there is a possible match. In theory this takes an infinite amount of time, but a solution to this problem can be found. The main issue comes from the fact that a metanotion can generate an infinite language  (i.e. an infinite number of words). What we want to do is to find the terminal productions of the metanotions which are in the left-hand side of the hyperrule so that, after substitution, it corresponds to the sentential form. So, we want to parse the sentential form in accordance to the \lq\lq{}metagrammar\rq\rq{}, with the left-hand side of the hyperrule as the starting form. When the parsing is done, we can deduce which are the terminal productions that have to be used to match the sentential form. As the metagrammar is a context-free language, it can be parsed efficiently. So the problem can be solved. Thus, with this mechanism a member is considered to be a terminal symbol if no match is found in the left-hand side of the hyperrules. So it is not needed to append the symbol \lq\lq{}symbol\rq\rq{} at the end of a member to make it a terminal symbol.

Now, we know how to produce words from a VW grammar. We know too that VW grammars can handle context-sensitivity. So now we want to write rules which transform one sentential form into another one, while preserving its semantic (its context's information).
In order to do so, we modified a little the mechanism of the grammar : the word we want to transform is used as the starting word, and we do not try to parse it. In fact, a sort of parsing is handled by the way the production process works. Moreover, we use a random generator during the production process, to enable the production to randomly generate any word of the language described by the grammar.
As an example take these \makebox{metarules :}
\begin{IEEEeqnarray*}{us}
N :: & 0; 1; 2;$\ldots$; 9; 0N; 1N;$\ldots$; 9N.\\
HEX :: & N; a; b; $\ldots$ ; f; a HEX; b HEX; \\
& $\ldots$; f HEX.\\
ADR :: & 0xN.\\
NUM :: & ADR; HEX.\\
INST :: & mov; push; pop.\\
REG :: & eax; ebx; edx.\\
STACK :: & esp.\\
REGS :: & STACK; REG.\\
REGNUM :: & REGS; NUM.\\
MEM :: & [ REGS ]; [ ADR ].\\
COMMA :: & \lq{},\rq{}.
\end{IEEEeqnarray*}
The metanotion $NUM$ represents an address or an hexadecimal number. The metanotion $INST$ represents three instructions (mov, push and pop). And so on.. \\
The hyperrules :
\begin{IEEEeqnarray*}{us}
mov REGS  CO&MMA REGNUM : \\
 & move REGNUM in REGS. \\
\IEEEeqnarraymulticol{2}{s}{push REGNUM :} \\
 & save REGNUM. \\
\IEEEeqnarraymulticol{2}{s}{pop REGS :} \\
 & restore REGS.
\end{IEEEeqnarray*}
modify an instruction into a readable sentence. For example the word \lq\lq{}mov eax, 0\rq\rq{} will be replaced by \lq\lq{}move 0 in eax\rq\rq{}, because of the first hyperrule.\\
We can add hyperrules which will transform these sentence into other equivalent sentence(s) :
\begin{IEEEeqnarray*}{ss}
move REG&NUM in MEM : \\
& mov, MEM, COMMA, REGNUM; \\
\IEEEeqnarraymulticol{2}{s}{move REGNUM in REGS :} \\
& mov, REGS, COMMA, REGNUM; \\
& save REGNUM, restore REGS. \\
\IEEEeqnarraymulticol{2}{s}{save REGNUM : push, REGNUM;} \\
& subtract 4 from esp, move \\
& REGNUM in [ esp ]. \\
\IEEEeqnarraymulticol{2}{s}{restore REGS : pop, REGS;} \\
& move [ esp ] in REGS, ADD 4 to esp. \\
& $\ldots$
\end{IEEEeqnarray*}

Now the sentential form obtained before (\lq\lq{}move 0 in eax\rq\rq{}) can be replaced by either \lq\lq{}mov, eax, \lq{},\rq{}, 0\rq\rq{} or by \lq\lq{}save 0, restore eax\rq\rq{}. If the first alternative is selected, then the generation will stop. Indeed, the sentential form is composed of \lq\lq{}mov\rq\rq{}, \lq\lq{}eax\rq\rq{}, \lq\lq{} \lq{},\rq{} \rq\rq{} and \lq\lq{}0\rq\rq{}, and none of these words match a left-hand side of a hyperrule. On the other side, if the second alternative is selected then the generation continues, and both parts of the sentential form, \lq\lq{}save 0\rq\rq{} and \lq\lq{}restore eax\rq\rq{}, can be replaced independently from each other. Thus, the sentential form \lq\lq{}save 0\rq\rq{} can be replaced by 
\lq\lq{}push, 0\rq\rq{} (so the generation stops) or by \lq\lq{}subtract 4 from esp, move 0 in [ esp ]\rq\rq{}, etc.

The metarules used above can be more sophisticated so they generate an infinite set of instructions, and so the hyperrules generate an infinite number of production rules. Hence we can have a (infinite) rewriting system handling an infinite number of instructions.

\section{\emph{K}-ary viruses}
\subsection{What is a \emph{K}-ary viruses}

\begin{definition}[\cite{DBLP:journals/virology/Filiol07}]
A \emph{K}-ary virus is composed of a family of \emph{k} files (some of which may not be executable), whose union constitutes a computer virus and performs an offensive action that is equivalent of that of a true virus. Such a code is said sequential if the \emph{k} constituent parts are acting strictly one after the another. It is said parallel if the \emph{k} parts executes simultaneously.
\end{definition}

The interest of combined virus lies in the fact that the viral information is split in various parts, which taken separately can have a non-malicious behaviour. Because of this separation of the viral information, we are out of the scope of Cohen's model. His model supposes that a virus is made of a unique sequence of symbols, which is not the case with combined viruses.

Two main classes of \emph{K}-ary viruses have been identified \cite{DBLP:journals/virology/Filiol07} :

\begin{itemize}
 \item Class 1 codes. These are the codes that work sequentially. \\
 This class is composed of three subclasses :
 \begin{itemize}
  \item Subclass A. Each code refers or contains a reference to the others. Thus, the detection of one of these codes leads to the detections of all of the others.
  \item Subclass B. None of the codes refers of contains a reference to the others. Thus, detecting one code does not affect the other codes. The detected code can be replaced by another code.
  \item Subclass C. The dependence of the code is directed. Thus detecting one code does not affect the codes which are before it in the sequential execution.
 \end{itemize}
 \item Class 2 codes. These are the codes that work in parallel.
 This class is composed of the same three subclasses as the class 1.
\end{itemize}

\subsection{Van Wijngaarden representation}

The power of a \emph{K}-ary virus lies in the fact that it is split in several parts. Thus, one can see a \emph{K}-ary virus as a distributed program whose global action is the same as that of a virus. If we look at this type of program from the point of view of formal grammars, we can feel that such a program can be described by them.

\begin{definition}
Let $\mathnormal{x_{1}, x_{2}}$ be two files, and $\mathnormal{v}$ $\in$ $\mathnormal{L(G_{v})}$ a virus. We define the relation $\mathnormal{R_{v}}$ by
\begin{center}$\mathnormal{x_{1} R_{v} x_{2} \Leftrightarrow \lbrace \exists \omega \in ( x_{1} \oplus x_{2} ) \mid \omega \in L(G_{v}) \rbrace}$ \end{center}
\end{definition}
The $\oplus$ operator is a selection function, whose result is a set of words over its input. The idea is that is does a selection of some parts of its inputs to extract a word from them, and if one of the results is in the language generated by $G_{v}$ then its inputs form a \emph{K}-ary virus.

The different parts of a \emph{K}-ary virus can each be described separately by a grammar. If we put all these parts together, we have the description of the virus as a whole. Thus a Van Wijngaarden grammar can be used to define \emph{K}-ary virus. The starting symbol produces all the parts of the \emph{K}-ary virus, then the different parts are recognized by some hyperrules of the grammar. The consistent substitution allows some informations to be shared between each parts while they are created. As an example, for a \emph{K}-ary virus with \emph{K}=3, the rules would look like :

\begin{IEEEeqnarray*}{us}
S : PAR&T1 INFOS, PART2 INFOS, PART3 INFOS\\
PART1 &INFOS : \\
VW-&Grammar of PART1 knowing INFOS\\
PART2 &INFOS : \\
VW-&Grammar of PART2 knowing INFOS\\
PART3 &INFOS : \\
VW-&Grammar of PART3 knowing INFOS\\
\end{IEEEeqnarray*}

Once the combined virus is produced (that is, that we have different files that contains the elements of the virus) each part may mutate on its own. While \emph{K}-ary malware have been formally defined \cite{DBLP:journals/virology/Filiol07} and their detection addressed, our approach enables to formalize the automatic generation of \emph{K}-ary malware while providing a constructive proof.

\section{Conclusion}

Van Wijngaarden grammars are very powerful, and can be easily understood by a human. The power of these grammars comes from the two context-free grammars that are jointly used, coupled to the uniform replacement rule which allows context-sensitive conditions to be expressed. 
It is thus possible to handle undecidable problems suitable to design undetectable malwarse in a far easier way than considering formal grammars of class 0 directly.

\emph{K}-ary virus have been defined through the use of a Van Wijngaarden grammar. The main idea is that the alternatives of the starting symbol are actually themselves the starting symbol of a grammar, describing each part (file) that the virus is composed of. This formal definition produces a constructive method to generate those codes automatically.

\section*{Acknowledgement}
The author would like to thank Eric Filiol for his fruitful discussions about formal grammars, his active support to this work and all the people at the Operational Cryptology and Virology lab for the wonderful, stimulating and friendly environment they generate.

\bibliographystyle{alpha}
\bibliography{all}

\end{document}